\documentclass{PoS}

\usepackage{epstopdf}

\title{One-loop electroweak corrections to $e^+e^-$ into three-jets}

\ShortTitle{One-loop electroweak Corrections to Three-jet observables}

\author{\speaker{Carlo Michel Carloni Calame}\\%
        INFN and University of Southampton\\
        E-mail: \email{c.carloni-calame@phys.soton.ac.uk}}
\author{Stefano Moretti\\
        University of Southampton\\
        E-mail: \email{stefano@phys.soton.ac.uk}}
\author{Fulvio Piccinini\\
        INFN, Sezione di Pavia\\
        E-mail: \email{fulvio.piccinini@pv.infn.it}}
\author{Douglas A. Ross\\
        University of Southampton\\
        E-mail: \email{dar@phys.soton.ac.uk}}

\abstract{We describe the impact of the full one-loop Electro-Weak  
terms of ${\cal O}(\alpha_{\rm S}\alpha_{\rm{EM}}^3)$ 
entering the electron-positron into three-jet cross-section.
We include both factorisable and non-factorisable virtual corrections,
photon bremsstrahlung but not the real emission of $W^\pm$
and $Z$ bosons. We show preliminary results and
we discuss the impact of the Electro-Weak
corrections on three-jet observables.}

\FullConference{8th International Symposium on Radiative Corrections \\
                 October 1-5, 2007\\
                 Florence, Italy}

\newcommand{\be}{\begin{equation}}
\newcommand{\ee}{\end{equation}}
\newcommand{\br}{\begin{eqnarray}}
\newcommand{\er}{\end{eqnarray}}
\newcommand{\ba}{\begin{array}}
\newcommand{\ea}{\end{array}}
\newcommand{\bi}{\begin{itemize}}
\newcommand{\ei}{\end{itemize}}
\newcommand{\bn}{\begin{enumerate}}
\newcommand{\en}{\end{enumerate}}
\newcommand{\bc}{\begin{center}}
\newcommand{\ec}{\end{center}}

\def\epem{\ifmmode{e^+ e^-} \else{$e^+ e^-$} \fi}

\newcommand{\Dir}{\kern -6.4pt\Big{/}}
\newcommand{\Dirin}{\kern -10.4pt\Big{/}\kern 4.4pt}
\newcommand{\DDir}{\kern -8.0pt\Big{/}}
\newcommand{\DGir}{\kern -6.0pt\Big{/}}

\def\jhep #1 #2 #3 {{JHEP} {\bf#1} (#2) #3}
\def\plb #1 #2 #3 {{Phys.~Lett.} {\bf B#1} (#2) #3}
\def\npb #1 #2 #3 {{Nucl.~Phys.} {\bf B#1} (#2) #3}
\def\epjc #1 #2 #3 {{Eur.~Phys.~J.} {\bf C#1} (#2) #3}
\def\zpc #1 #2 #3 {{Z.~Phys.} {\bf C#1} (#2) #3}
\def\jpg #1 #2 #3 {{J.~Phys.} {\bf G#1} (#2) #3}
\def\prd #1 #2 #3 {{Phys.~Rev.} {\bf D#1} (#2) #3}
\def\prep #1 #2 #3 {{Phys.~Rep.} {\bf#1} (#2) #3}
\def\prl #1 #2 #3 {{Phys.~Rev.~Lett.} {\bf#1} (#2) #3}
\def\mpl #1 #2 #3 {{Mod.~Phys.~Lett.} {\bf#1} (#2) #3}
\def\rmp #1 #2 #3 {{Rev. Mod. Phys.} {\bf#1} (#2) #3}
\def\cpc #1 #2 #3 {{Comp. Phys. Commun.} {\bf#1} (#2) #3}
\def\sjnp #1 #2 #3 {{Sov. J. Nucl. Phys.} {\bf#1} (#2) #3}
\def\xx #1 #2 #3 {{\bf#1}, (#2) #3}
\def\hepph #1 {{\tt hep-ph/#1}}

\def\beq{\begin{equation}}
\def\beeq{\begin{eqnarray}}
\def\eeq{\end{equation}}
\def\eeeq{\end{eqnarray}}
\def\a0{\bar\alpha_0}

\def\b0{\beta_0}

\def\ee{e^+e^-}

\def\slashchar#1{\setbox0=\hbox{$#1$}           
     \dimen0=\wd0                                 
     \setbox1=\hbox{/} \dimen1=\wd1               
     \ifdim\dimen0>\dimen1                        
        \rlap{\hbox to \dimen0{\hfil/\hfil}}      
        #1                                        
     \else                                        
        \rlap{\hbox to \dimen1{\hfil$#1$\hfil}}   
        /                                         
     \fi}                                         %

\def\be{\begin{equation}}
\def\ee{\end{equation}}
\def\bea{\begin{eqnarray}}
\def\eea{\end{eqnarray}}

\def\slash{/\kern -5pt}

\def\ims #1 {\ensuremath{M^2_{[#1]}}}

\def\s22w{s_{2W}^2}

\begin{document}
\section{Three-jet Events at Leptonic Colliders}
\label{Sec:Intro}
Strong (QCD) and Electro-Weak (EW) interactions
are two fundamental forces of 
Nature, the latter in turn unifying
Electro-Magnetic (EM) and Weak interactions
in the Standard Model (SM).
A clear hierarchy exists between the strength of these two interactions
at the energy scales probed by past and present high energy particle 
accelerators (e.g., LEP, SLC, HERA and Tevatron): 
QCD forces are stronger than EW ones.
This argument, however, is only valid in lowest order
in perturbation theory. 

A peculiar feature in fact distinguishing QCD and EW effects in higher orders
is that the latter are enhanced by double logarithmic factors,
$\log^2(\frac{s}{M^2_{{W}}})$, 
which, unlike in the former, do not cancel for `infrared-safe' 
observables \cite{Kuroda:1991wn,Beenakker:1993tt,Ciafaloni:1999xg}. 
The origin of these `double logs' is well understood.
It is due to a lack of the Kinoshita-Lee-Nauenberg (KLN) 
\cite{KLN} type 
cancellations of Infra-Red (IR) -- both soft
and collinear -- virtual and real emission in
higher order contributions originating from $W^\pm$ (and, possibly,
$Z$) exchange. 
This is in turn a consequence of the 
violation of the Bloch-Nordsieck theorem \cite{BN} in non-Abelian theories
\cite{Ciafaloni:2000df}.
The problem is in principle present also in QCD. In practice, however, 
it has no observable consequences, because of the final averaging of the 
colour degrees of freedom of partons, forced by their confinement
into colourless hadrons. This does not occur in the EW case,
where the initial state has a non-Abelian charge,
dictated by the given collider beam configuration, such as in $e^+e^-$
collisions. 

These logarithmic corrections are finite (unlike in
QCD), as the masses of the weak gauge bosons provide a physical
cut-off for $W^\pm$ and $Z$ emission. Hence, for typical experimental
resolutions, softly and collinearly emitted weak bosons need not be included
in the production cross-section and one can restrict oneself to the 
calculation of weak effects originating from virtual corrections and
affecting a purely hadronic final state. Besides, these contributions can  be
isolated in a gauge-invariant manner from EM effects
\cite{Ciafaloni:1999xg}, 
at least in some specific cases, and 
therefore may or may not
be included in the calculation, depending on the observable being studied. 
As for purely EM effects,
since
 the (infinite) IR real photon emission cannot be resolved experimentally, 
this ought to be combined with the (also infinite) virtual one, through the 
same order, to recover a finite result, which is however not
doubly logarithmically enhanced (as QED is an Abelian theory).
 

In view of all this,
our aim is the computation of the full 
one-loop EW effects entering three-jet production in electron-positron
annihilation at any collider energy
via the subprocesses $e^+e^-\to\gamma^*,Z\to \bar 
qqg$\footnote{See Ref.~\cite{2jet} for the corresponding one-loop
corrections
to the Born process $e^+e^-\to\bar qq$ and Ref.~\cite{4jet} for the
$\sim n_{\rm f}$ component of those to $e^+e^-\to \bar qqgg$ (where 
$n_{\rm f}$ represents the number of light flavours).}.
Ref.~\cite{oldpapers} tackled part of these, in fact, limitedly to the
case of $W^\pm$ and $Z$ (but not $\gamma$) exchange and 
when the higher order effects arise only from initial or final state
interactions 
(these represent the so-called `factorisable' corrections, i.e.,
those involving loops 
not connecting the initial leptons to the final quarks,
which are the dominant ones at $\sqrt s=M_Z$, where the width 
of the $Z$ resonance provides a natural cut-off for off-shellness
effects). The remainder, `non-factorisable' corrections,
while being typically small at $\sqrt s=M_{Z}$, 
are expected to play a quantitatively relevant role as $\sqrt s$ grows
larger.  By studying the full set
of the one-loop EW corrections, we improve on the
results of Ref.~\cite{oldpapers} in two respects: (i) we include now all
the non-factorisable terms; (ii) we also incorporate previously
neglected genuine QED corrections, including photon bremsstrahlung.

Combining the logarithmic enhancement associated to
the genuinely weak component of the EW corrections to the fact that
$\alpha_{\rm S}$ steadily decreases with energy, 
unlike $\alpha_{\rm{EM}}/\sin^2\theta_W$,    
in general, one expects 
one-loop EW effects to become comparable to QCD ones
 at future Linear Colliders
(LCs) \cite{LCs} running at TeV energy scales\footnote{For example, 
at one-loop level,
in the case of the inclusive cross-section of $e^+e^-$
into hadrons, the QCD corrections are of  ${\cal O}
(\frac{\alpha_{\mathrm{S}}}{\pi})$, whereas
the EW ones are of ${\cal O}(\frac{\alpha_{\mathrm{EW}}}{4\pi}\log^2
\frac{s}{M^2_{{W}}})$, where $s$ is the collider CM energy
squared, so that at $\sqrt s\approx1.5$ TeV the former are identical to the latter,
of order 9\% or so.}. In contrast, 
at the $Z$ mass peak, where logarithmic
enhancements are not effective, one-loop EW corrections are expected to appear
at the percent level, hence being of limited relevance at
LEP1 and SLC, where the final error on $\alpha_{\mathrm{S}}$
is of the same order or larger \cite{Dissertori}, but of crucial importance
at a GigaZ stage of a future LC \cite{oldpapers}, where the relative accuracy
of $\alpha_{\mathrm{S}}$ measurements is expected to be at the
$0.1\%$ level or smaller \cite{Winter}.
On the subject of higher order QCD
effects, it should be mentioned here that a great deal of effort has    
recently been devoted to evaluate two-loop contributions
to the three-jet process \cite{QCD2Loops}
while the one-loop QCD results have been known for quite some time \cite{ERT}.

As intimated, in the case of $e^+e^-$ annihilations, the most important QCD quantity to be 
extracted from multi-jet events is $\alpha_{\mathrm{S}}$.
The confrontation of the measured value of the strong coupling
constant with that predicted by the theory through the 
renormalisation group evolution is an important test of the 
SM or else an indication of new physics, when its typical mass scale is 
larger than the collider energy, so that the new particles cannot be
produced as `real' detectable states
but may manifest themselves through `virtual' effects. 
Not only jet rates,
but also
jet shape observables would be affected.
\begin{figure}
\begin{center}
\includegraphics[width=12cm]{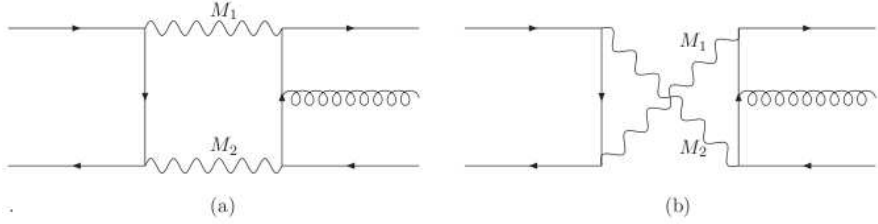}
\end{center}
\caption{Pentagon graphs. The gauge bosons in the loop can be $W$,
$Z$ or $\gamma$.}
\label{fig:pentagons}
\end{figure}
Our calculation involves the full
one-loop EW corrections
to three-jet observables in electron-positron annihilations,
including also non-factorisable corrections such as the ones
generated via the interference of the pentagon graphs
in Fig.~\ref{fig:pentagons} with the tree-level 
ones. Hence, our calculation not
only accounts for the mentioned double logarithms, but also all single ones
as well as the finite terms arising through the complete 
 ${\cal O}(\alpha_{\rm S}\alpha_{\rm{EW}}^3)$. We account for all
possible flavours of (anti)quarks in the final state, with the exception
of the top quark. The latter however appears in some of the loops whenever
a $b\bar bg$ final state is considered, in particular notice that, in this
case, we will also have to include loops involving the Higgs boson coupling 
to (anti)top quark lines. 

We expect that all such corrections are of a few percent at 
$\sqrt s=M_Z$ and that they grow to a few tens of percent
at LC energies. Hence, while their impact is not dramatic in the context
of LEP1 and SLC physics at a GigaZ stage of future LCs
they ought to be taken into account in the experimental fits.
Even more so, it is the case of future LCs running at and beyond
the TeV range.

%

\section{Calculation and preliminary results}
\label{concl}
\begin{figure}
\begin{center}
\includegraphics[width=7.6cm]{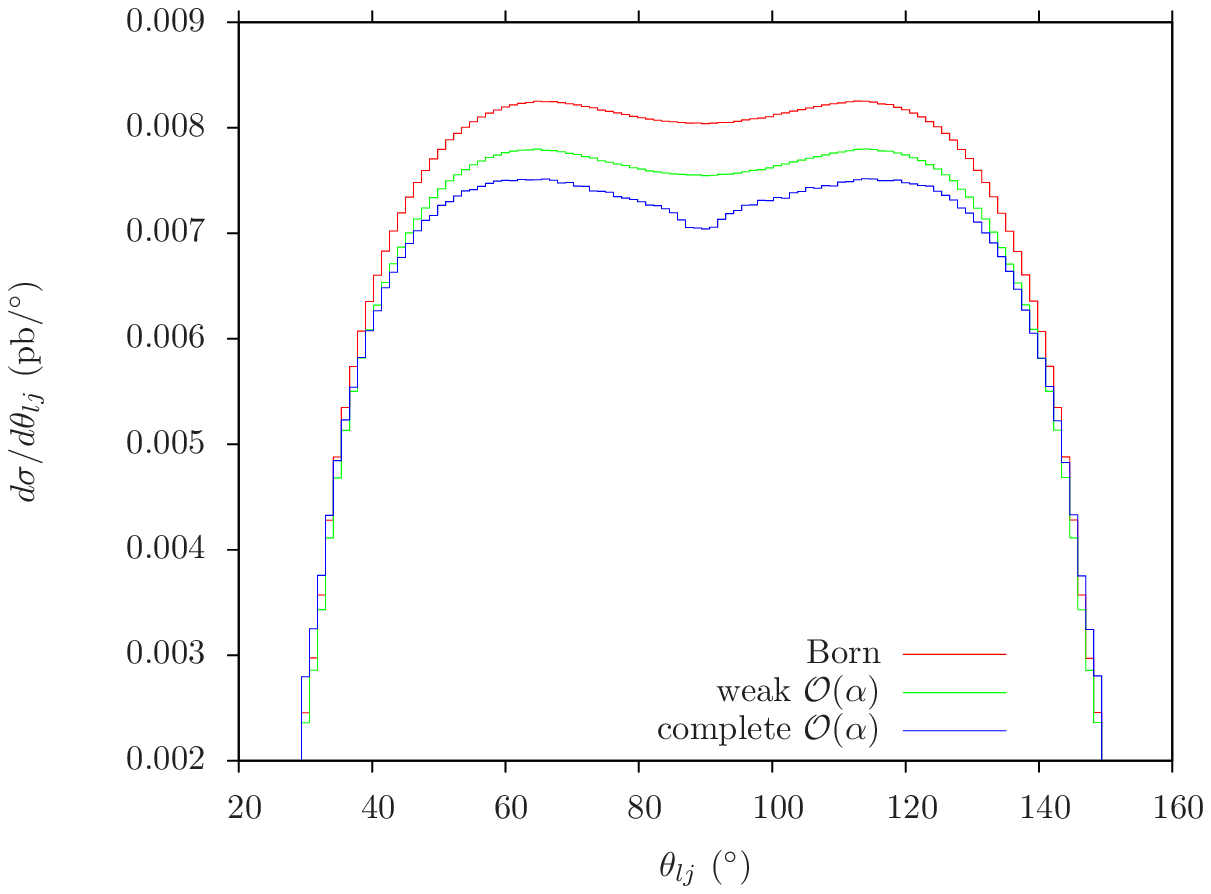}~\includegraphics[width=7.6cm]{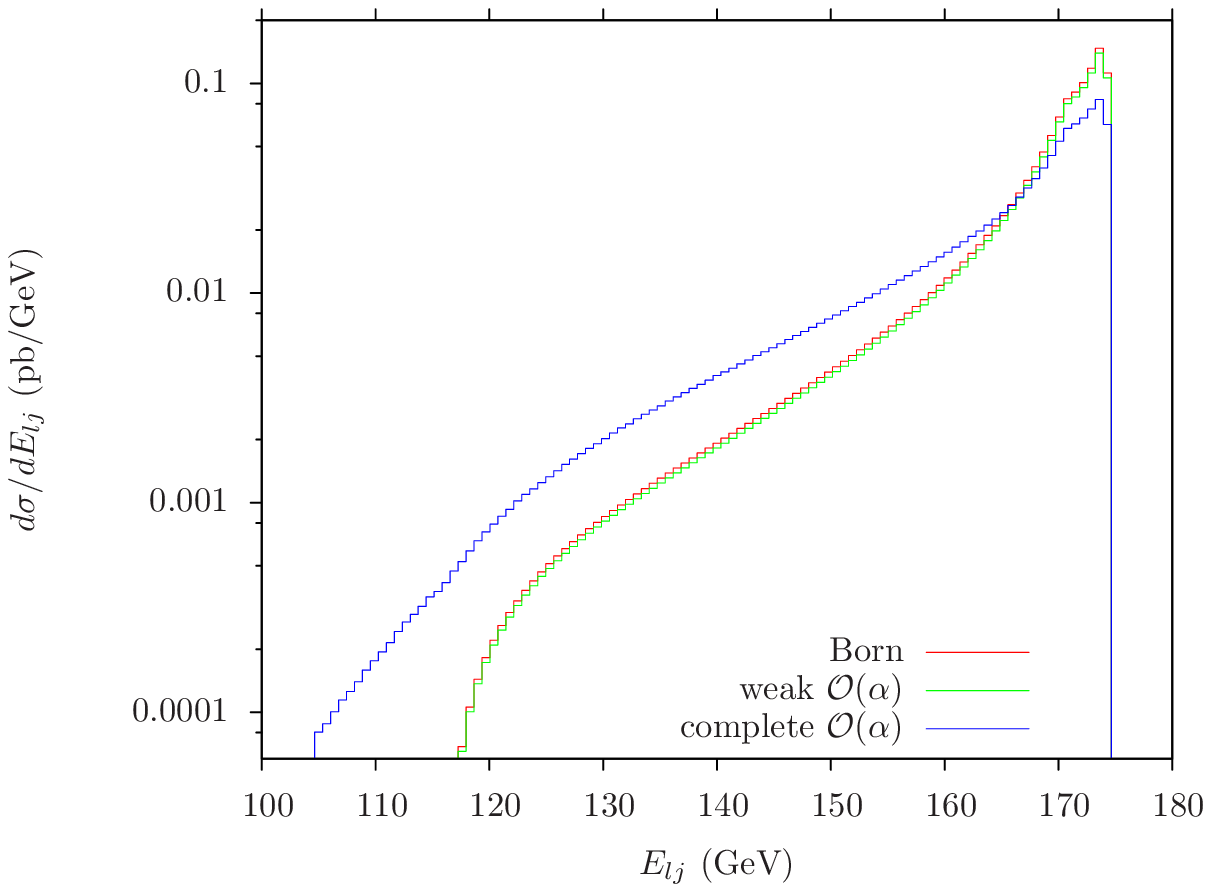}
\end{center}
\caption{Distributions of the leading jet angle and energy}
\label{fig:ajlejl}
\end{figure}
\begin{figure}
\begin{center}
\includegraphics[width=7.6cm]{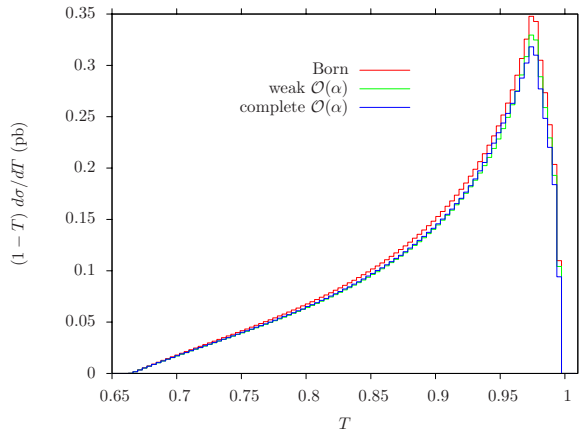}
\end{center}
\caption{Thrust distribution}
\label{fig:thrust}
\end{figure}
In this section we sketch the main features of the calculation and
we present preliminary results. The detailed description of the
calculation and a wider phenomenological study appeared
elsewhere~\cite{CMPRinpreparation}.

Since with respect to Refs.~\cite{oldpapers} we include QED
corrections, loop diagrams can contain one or two photons and give rise to
infrared (IR) and collinear divergences.
We regularise the divergences by simply inserting a mass $\lambda$ for
the photon and
a mass $m_f$ for all fermions. This is also done in the case of
the bremsstrahlung contribution before integrating over the 
phase space for the emitted photon. In order to check the
cancellation of the IR divergences between real and virtual
corrections, we successfully verified the independence of their sum
against variation of the photon mass $\lambda$.
Another key feature of this calculation 
is the
occurrence of pentagon graphs, as shown in
Fig.~\ref{fig:pentagons}. Such graphs 
involve five-point Passarino-Veltman (PV)~\cite{PV} functions 
with up to three powers of momenta in the numerator. We have 
handled these in two 
separate ways (with two independent codes), in order 
to check for possible numerical instabilities. In the first case
the integrals are simply 
evaluated using routines in LoopTools 
v2.2~\cite{looptools}\footnote{We implemented also 
directly in an independent fortran routine the expressions for the 
five point functions of Ref.~\cite{DD}, 
finding agreement with LoopTools.}. In the other 
we use the standard PV
reduction, 
carried out exhaustively until only scalar pentagon
integrals appear. The latter are available in the library
FF1.9~\cite{FF1.9}. A comparison of  the numerical results
provided by the two codes 
yielded satisfactory agreement between the two 
methods.
Also the squared amplitudes for the real emission process have been
evaluated by using two independent tools (ALPHA~\cite{alpha} and
MadGraph~\cite{madgraph}) finding perfect agreement. The
integration over the three- and four-body phase space is performed
numerically by means of a Monte Carlo method, using standard
importance sampling techniques for the variance reduction.

In order to perform a preliminary analysis, we considered an
$e^+e^-$ collider at $\sqrt{s}=350$~GeV and we used a realistic
experimental setup: partonic momenta are clustered into jets 
according to the
Durham jet algorithm~\cite{durham} (e.g. when $y_{ij}<y_{min}$
with $y_{min}=0.001$), the jets are required to lie in the central
detector region $30^\circ<\theta_{\mathrm{jets}}<150^\circ$ and we require that
the invariant mass of the jet system is larger than $0.75\times\sqrt{s}$.
If a real photon is present in the final state, it is
clustered according to the same algorithm, but we require that
at least 3 ``hadronic'' jets are left at the end. Finally, we sum
over the final-state quarks.

In Figs.~\ref{fig:ajlejl} and~\ref{fig:thrust}, some examples of
the impact of the full EW corrections on three-jet observables are
shown. The distributions
are plotted in Born approximation (red line), including the complete
1-loop EW corrections (blue) and including only the weak
corrections (green).
In Fig.~\ref{fig:ajlejl} the distributions of the
angle (left) of the most energetic (leading) jet and the energy 
(right) of the leading jet are shown. In Fig.~\ref{fig:thrust} the
distribution of the thrust shape variable is plotted.

Even in this limited set of observables, the impact of the EW
corrections is evident. The leading jet angle and energy distributions 
show clearly the effect of the purely weak and QED corrections.
The 5-6\% effect on the thrust
distribution indicates that the EW corrections are unavoidable to carry out
a precise measurement of $\alpha_s$ at a future LC with a 0.1\% accuracy. 
A more complete set of observables
are discussed in detail in
the paper~\cite{CMPRinpreparation}.

Before concluding, it is worth noticing that this calculation can be
used as a starting point (by exploiting the crossing symmetry) to
calculate the complete 1-loop EW corrections to
$\gamma^\star/Z +$~jet production at hadron colliders, which is a
process of great interest for physics at the forthcoming LHC. This
study is under consideration.

C.M.C.C. wants to thank the organizers for the
the stimulating atmosphere during the conference and
its success.


\end{document}